\documentstyle[12pt,aaspp4]{article}
\newcommand{\T}{\mathop{\rm T}\nolimits}
\begin{document}

\hfill Submitted to The Astronomy Letters.

\title{Activity of the Soft Gamma Repeater SGR 1900+14 in 1998 
from Konus-Wind Observations:\\
 2. The Giant August 27 Outburst.}

\author{E.P.Mazets\altaffilmark{1},
        T.L.Cline\altaffilmark{2},
        R.L.Aptekar\altaffilmark{1},
        P.Butterworth\altaffilmark{2},
        D.D.Frederiks\altaffilmark{1},
        S.V.Golenetskii\altaffilmark{1},
        V.N.Il'inskii\altaffilmark{1},
        V.D.Pal'shin\altaffilmark{1}
        }

\altaffiltext{1}{Ioffe Physical-Technical Institute, St.Petersburg, 194021, Russia. Mazets@pop.ioffe.rssi.ru}

\altaffiltext{2}{Goddard Space Flight Center, Greenbelt, MD 20771, USA.}

\vspace{2cm}

\begin{abstract}

The results of observations of the giant 1998 August 27 outburst 
in SGR 
1900+14 are presented. A comparison is made of the two 
extremely intense  
events on August 27, 1998 and March 5, 1979. The 
striking similarity between the outbursts strongly implies a common 
nature. The observation of two giant outbursts within 20~years from different 
sources  suggests that such events  occur 
in an SGR once every 50--100~years.

\end{abstract}

\clearpage

\section{INTRODUCTION}

The discovery of soft recurrent bursts started with the 
observation of the famous gamma-ray outburst on March 5, 1979 
(Mazets et al., 1979a). On the next day, the Konus experiment on the 
Venera~11 and 12 spacecraft detected the first recurrent burst 
arriving from the same source (Mazets et al., 1979a). The number of detected 
recurrent bursts increased subsequently to 16 (Golenetskii et al., 1984). At the 
end of March 1979, the Konus experiment discovered and localized 
a second recurrent burster, B1900+14 (Mazets et al., 1979b), and it is the new 
reactivation of the latter that is being considered in the 
present paper.

Recurrent bursts have been observed from these and more recently 
discovered soft gamma repeaters during rare 
periods of activity separated by long quiet intervals. 
However the March~5 event remained 
until recently the only one of its kind. Accordingly, the 
question of the extent to which the 1979 March~5 event is 
characteristic of the activity of soft gamma repeaters, of 
whether such events are typical for all SGRs, or this 
giant outburst was a unique episode in the history of one object 
only, remained unclear. Arguments for both the first (Mazets et al., 1982) and 
second (Norris et al., 1991; Fenimore et al., 1996) alternative were put forward.

An unambiguous answer to this question was obtained on August~27, 
1998, when several spaceborne experiments detected a giant 
outburst in SGR 1900+14, which was strikingly similar to the one 
that had occurred in SGR~0526-66 on 1979, March 5 (Cline et al., 1998; 
Hurley et al., 1999a; Feroci et al., 1999).

In this paper, we are going to consider in detail the results of 
the Konus-Wind observations of the 1998 August~27 outburst and to 
compare it with the 1979 March~5 event.

\section{OBSERVATIONS}

The outburst on August 27, 1998 has no precedents in its 
intensity. 
The peak flux of hard photons with energies $>15$~keV is 
considerably in excess of the level observed heretofore from any 
known cosmic sources. The overall time history of the 
burst is displayed in Fig.~1. As in the March~5 event, the burst 
starts with a narrow radiation pulse, which falls off rapidly to 
become a slowly decaying, coherently pulsating tail. 

We will consider in detail this and other characteristic stages in the 
development of the event.

\subsection*{a) Initial pulse and the $\T-\T_0<1$~s region}

The Konus-Wind cosmic gamma-ray burst spectrometer (Aptekar et al., 1995)
starts a 
program of detailed measurements from the instant $\T_0$ at which 
the trigger signaling the arrival of a burst is generated. This 
program includes measurement of the burst time history in three 
energy windows G1 (15--50~keV), G2 (50--250~keV), and
G3 (250--1000~keV) with a time resolution of 2 to 256~ms. The burst 
prehistory, i.e. some measurements preceding $\T_0$, is also 
stored. The trigger signal is generated in the second energy 
window G2 (50--250~keV) after the count rate has risen in a short 
time by $\sim 7$ standard deviations. The $\T_0$ signal 
also initiates measurements of multichannel energy spectra with automatic 
adaptation of the accumulation time to the current count rate. 

Figure 2 presents the initial part of the burst profile 
recorded within $\T-\T_0$ from $-0.5$ to 1~s. A weak precursor is 
observed in the soft-energy window 0.45~s before $\T_0$. As seen 
from the graph, the number of additional counts required to 
generate the trigger arrives in as short a time as 4~ms, so that 
the intensity rises very rapidly. In another 4~ms, no counts are 
detected in any window. This implies that the radiation intensity 
becomes so high only 4~ms after $\T_0$ that the count-rate channel 
completely overloads.

The instrument remains silent for a time, to revive 
again after $\sim 200$~ms. The dead time and pulse 
pile-up effects in the detector distort the real picture so 
strongly as to make correcting the data obtained an extremely 
difficult problem. To solve it, thorough laboratory studies of 
counting losses and pulse superposition effects were carried out 
on the spare hardware of the instrument at incident fluxes of 
$10^6{-}10^7$~photons~s$^{-1}$ and higher. The measurements were 
performed with radioactive sources and X-ray equipment, whose 
radiation permitted a satisfactory simulation of 
the required continuous spectra. A numerical model of the instrument 
response to high radiation fluxes was also made. Because of the 
importance of the data obtained for a correct evaluation of the 
event energetics, we will discuss them in detail. 

Under normal conditions and with most devices, the total photon flux  
$N$ incident on a detector is related to the recorded count rate 
$n$ through the obvious expression $1/n-1/N=\tau$, where 
$\tau$ is the instrument dead time. For $N\gg 1/\tau$, $n$ should 
tend to the limiting value $n=1/\tau$ and become independent of 
$N$. Figure~3 exemplifies the results of both laboratory 
measurements and numerical simulation of the count rates in the 
G1, G2 and G3 windows, and of their sum, as functions of input 
intensity for a spectrum representing a broad line with an energy 
$\sim 100$~keV. As can be seen   from the graphs, the standard $n(N)$ 
relationship holds until $N\sim 10^6$, but as the intensity $N$ 
continues to increase, the detected number of counts $n$ drops 
dramatically to finally vanish altogether. This behavior has a 
straightforward explanation. For high $N$, overlap of individual 
pulses in the detector gives rise to the formation and growth of 
a dc electric current component. Single-photon counts start to 
be replaced by detection of output-signal fluctuations, which 
causes extremely strong distortions of the measured 
spectrum. Indeed, the count rate in the third, hard window 
displayed in Fig.~3 is totally due to scintilattion 
pile-up in the detector. As the input flux continues to 
increase, the growth of the fluctuating dc signal at the 
amplifier output becomes limited by the power supply voltage. 
After this saturation level has been reached, the count rate 
drops dramatically, to vanish altogether when it is exceeded. 
Laboratory studies also revealed the important roles played by two 
aftereffects, which must be taken into account when the 
intensity undergoes sharp variations near the saturation level. 
The first of these is associated with the afterglow of the
long-lived phosphorescence component in the NaI(Tl) crystal, which 
results in time profile tailing. The second effect is connected 
with the slow response of the PMT high-voltage stabilization 
circuit. As a result, following large intensity jumps near 
shut-down the photomultiplier gain can deviate 
from the nominal value for 30--40~ms. It is this instrumental effect reproduced 
in laboratory conditions that accounts for the short-period 
appearance of counts during the interval $\T-\T_0=20-30$~ms.

Note the good agreement between direct measurements and numerical 
simulation of the $n(N)$ relations. This is a very important 
point. The pattern of the $n(N)$ relations for each of the three energy 
windows G1, G2, and G3 turns out to be strongly sensitive 
to the shape of the incident energy spectrum. Hence 
by comparing the observed count rates $n_1$, $n_2$, and $n_3$ one 
can reliably estimate not only the incident intensity but 
the spectral hardness as well. The possibility of spectral shape 
selection in laboratory conditions was limited, but the agreement 
between the results of direct measurements and simulation permits 
one to apply the simulation of the $n(N)$ relations to 
spectra of any shape.

Turning back now to the August~27 event and Fig.~1, we may 
conclude that while the problem of reproducing the shape of the 
burst profile during the first second after $\T_0$ is very 
difficult, it can be solved, except for the intervals of total 
shut-down. We have also been able to obtain important 
limiting estimates for the $\T-\T_0=0-0.2$~s interval. 
Note the following two important points. First, during the  
last few ms of the prehistory before $\T_0$ the count rates are still 
not high enough to become distorted. Their ratio in the three 
energy windows permits one to estimate the spectral hardness at 
the very beginning of the burst, which was found to be very high, 
corresponding for a $\propto E^{-1}\exp(-E/kT)$ spectrum 
to $kT\sim 200$~keV. Second, close to saturation the detector 
starts to operate as a calorimeter. The dc component of the PMT 
output current is proportional to the average photocathode 
illumination by scintillations, i.e. to the total energy released 
in the detector per unit time. The output current corresponding 
to saturation for a given spectrometer gain is known. Numerical 
simulation permits one to also take into account the fluctuations 
in the PMT output current, whose level depends on the hardness of 
the incident photon spectrum. Hence we can determine with a high 
accuracy the incident energy flux on the detector corresponding 
to complete suppression of counting. For instance, for incident 
photon spectra with $kT=30$ and 300~keV it was found to be 
$2.4\times 10^{-2}$ and $3.1\times 10^{-2}$~erg~cm$^{-2}$~s$^{-1}$, 
respectively. Thus the lower limit of the burst energy flux at 
total shut-down is established reliably.

Interestingly, total overload of the Konus-Wind detectors has been
observed to occur more than once in so-called imitation 
bursts, which are produced by ultrarelativistic multi-charged 
cosmic-ray nuclei interacting with the NaI(Tl) crystal. An 
instantaneous release of a tremendously high energy in the 
crystal is accompanied by a slowly decaying afterglow of the 
long-lived phosphorescence component with decay 
time $\tau \sim 100-150$~ms (Koi\v{c}ki S., Koi\v{c}ki A., Ajda\v{c}i\'c, 1973).
The PMT output current 
fluctuations are detected as individual X-ray photons. Figure~4 
illustrates a moderately strong imitation burst observed 
on July~22, 1997. The output current 
due to the particle energy loss instantly overloads the amplifier. 
After the 
current has decayed below the saturation level, an enormous count 
rate appears in the G1 window. This example (compare with 
Fig.~2) also reveals a short-lived appearance of counts due 
to the slow response of the PMT power-supply stabilization 
system. Figures~2 and 4 differ radically in the 
count-rate levels in the G2 and G3 hard 
windows. This results from differences in the amplitude of 
fluctuations in the output current and their scatter, caused by 
the fact that the saturation in the August~27 event occurs as a 
result of superposition of narrow scintillation trains caused by 
hard gamma photons, whereas the imitation burst is actually a sum 
of an enormous number of small signals produced by individual 
phosphorescence photons.

Within the $\T-\T_0=0.2-1$~s interval, the photon flux incident on 
the detector is still very high. It undergoes strong and sharp 
variations both in intensity and spectral composition. Four 
detector spectra were obtained within this interval 
with a 0.256~s accumulation time. Such averaging is too coarse to allow 
determination of fast changes in the intensity and hardness. 
Nevertheless, Fig.~5 shows convincingly that on the 
average the burst radiation hardness within the $\T-\T_0=0.512-0.768$~s
interval is substantially higher. 

Numerical simulation offers the possibility of reconstructing the 
initial levels of intensity and hardness with a high temporal 
resolution. We are speaking about reconstruction 
rather than introduction of corrections, because the measured and 
real intensities may differ by orders of magnitude. The procedure of 
reconstruction consists essentially in finding for each three 
count rates in the three windows $n_i$ (see Fig. 2) the 
pair of values of $N$ and $kT$ such that the calculated values of 
$n_i(N,kT)$ provide the best fit to the measurements. One has to 
understand by the $kT$ parameter here the hardness characteristic 
of the burst energy spectrum adopted in the calculations for a 
given instant of time. The spectral shape used most frequently is 
that of thermal bremsstrahlung radiation.
Power-law and Band~(1993) spectra were also tried.

The first second of the burst time history reconstructed using this 
procedure is displayed in Fig.~6. Also shown is the 
variation of $kT$. We readily see that the initial pulse decays 
nonmonotonically. The smoothness of the exponential falloff 
($\tau \sim 35$~ms) is broken by two additional peaks in 
intensity. The radiation spectrum varies very strongly. In the 
very beginning of the pulse ($\T-\T_0=0-4$~ms), $kT\sim 300$~keV. The 
radiation in the decaying part turns out to be very soft, 
$kT \sim 20$~keV. By the time $\T-\T_0 \sim 350$~ms, the radiation intensity 
drops down to $\sim 5\times 10^5$~photons~s$^{-1}$. The $kT$ remains 
low, $\sim 20$~keV. After this the intensity begins to grow again 
to increase by an order of magnitude by the time $\T-\T_0=550$~ms. 
The temperature increases to 250~keV. Finally, the intensity 
falls off to become a long, slowly decaying tail. The dashed line 
at the top of the initial pulse specifies the lower intensity 
limit corresponding to instrument shut-down. The real intensity in the 
initial pulse considerably exceeded 
this limit, possibly by a large factor. The initial 
pulse apparently ends by the time $\T-\T_0 \sim 300$~ms. The new, 
nearly tenfold increase in intensity and, most significantly, the 
sharp growth in radiation hardness possibly indicate a 
manifestation of some new aspects of the burst process and, 
perhaps, of new additional sources of energy.

Thus the first second in the outburst time history appears to have 
been very complex. It should be noted that information on the 
initial phase of the August~27 burst observed from Ulysses 
(Hurley et al., 1999a) and BeppoSAX (Feroci et al., 1999) was irretrievably 
lost.

\subsection*{b) Transition region, $\T-\T_0=1-35$~s, and subsequent 
pulsations} 

Within this time interval (Fig.~7), the intensity falls 
off slowly. The monotonic course of the decay is broken by fairly 
strong variations revealing a periodicity with $P=5.16$~s. The 
observed count-rate level allows straightforward dead-time 
corrections. Figure~7 also shows the variation of the 
count-rate ratio in two energy windows, which characterizes 
spectral hardness. The spectra obtained at the beginning and at 
the end of this interval are illustrated by Fig.~8.

Starting from the 35th second, the strict periodicity of the 
radiation becomes increasingly distinct (Fig.~7). The 
pulsation amplitude increases strongly. The pattern of pulsations 
does not remain stable. In the beginning of this region, the 
$P=5.16$~s period is represented by four separate peaks. The peaks 
decrease gradually in relative amplitude. By the end of the 
observations, they become barely visible, but the modulation at 
the fundamental frequency is still seen. The hardness curve 
likewise behaves in a very complex manner. Averaged over 
of the period, the 
hardness remains practically constant. This is corroborated by 
the spectral measurements (Fig.~8), which yield 
values of $kT$ within $23 \pm 3$~keV, but the pulsations are 
accompanied by small-scale hardness variations. During the first 
100~s of the burst, the hardness clearly correlates with the 
intensity of individual peaks. Nearer the end of the burst, the 
situation reverses. The hardness of the radiation component 
modulated at the main frequency of 0.194~Hz is observed to be 
highest at the minima of the intensity.

\section{DISCUSSION}

The two giant outbursts, observed on March~5, 1979 in SGR~0526-66 
and on August~27, 1998 in SGR~1900+14, were strikingly similar. 
Combined analysis of these events is of considerable interest. 
Therefore for the sake of convenience we present in Fig.~9 
the 1979 March~5 burst profile as observed 
from the Venera~11 and 12 spacecraft, and in Fig.~10, the time history of the 
1998 August~27 event. Each event consists of a short, 
superintense radiation pulse followed by an exponentially 
decaying tail. The tail exhibits, in its turn, strictly periodic 
pulsations, which set in and develop rapidly. Immediately after 
the initial pulse one observes an intensity increase in the form 
of a single wave, which is not in phase with the train of 
periodic pulsations. Fairly fast irregular oscillations appear at 
the crest of the wave. In both cases the initial pulse 
is $\sim 0.25$~s long. Most of the intensity increase occurs during a 
short time of $\sim 2$~ms. By contrast, in the August~27 event 
one observes, besides a weak precursor at the very beginning of 
the pulse, an 80-ms long slow increase in intensity, and it is 
this increase that transforms to the steep rise. If such details 
did exist in the March~5 event, they were below the detection 
threshold. In this event as seen from Fig.~9, the initially steep rise 
is replaced by a slower increase of the intensity. Accordingly, 
the maximum in the pulse is reached in 50--100~ms. 
One may conjecture that the initial pulse in the August~27 event likewise 
increased with a slowing down of the rise,
judging from 
the short-time observation of counts within the $\T-\T_0=30-40$~ms 
interval (Fig.~2). We recall that such an effect is possible if the 
intensity slightly exceeds the saturation level.

In both events, the decay of the initial pulse is close to 
exponential with a time constant $\sim 40$~ms. An additional hump 
is observed in the intensity profile of each event after the 
initial pulse. In contrast to SGR~0526-66, the distinct recurrent 
pattern of periodic pulsations in SGR~1900+14 does not set in 
immediately, but after 35--40~s. However during this interval the 
power spectrum likewise exhibits a clearly pronounced peak at a 
frequency of 0.194~Hz.

The pulsating tails in both events decay with exponential constants of 
80--100~s. The behavior of the energy spectra also exhibits a 
pronounced similarity. The spectrum of the initial pulse in the 
SGR~0526-66 burst contains a hard component with an emission 
feature around 400~keV (Mazets et al., 1979a). The pulsations have a soft 
spectrum with $kT=30-35$~keV, similar to those of the later 
bursts. Because of the giant intensity of the initial pulse, 
accurate spectral data for the August~27 event could not be 
obtained. It has, however, been reliably established that the 
spectrum at the very beginning of the pulse after $\T_0$ is very hard, 
$kT\sim 300$~keV, to become very soft, with $kT \sim 20$~keV, in 
its tail, below the saturation limit. A similar fast spectral 
evolution during the development and decay of the initial pulse 
could possibly have occurred in the March~5 event too, and it is 
the fast spectral variability that accounts for the presence of a 
hard and a soft component in the resultant spectrum.

It is appropriate to note here that the spectrum of the initial 
pulse in the March~5 event measured on the Venera spacecraft 
(Mazets et al., 1979a) was questioned by Fenimore et al~(1996). 
Unfortunately, those authors based their consideration on a wrong
assumption that the dead 
time in the multichannel 
spectral measurements was $\sim 1$~ms and that, accordingly, the 
number of counts accumulated during $\sim 200$~ms was very small, 
not more than 200 for the whole spectrum. In fact,
the 1~ms-time was the dead time under independent accumulation of 
counts in each of the 16 spectrometer channels separately, a 
condition providing reliable spectral measurements.

The second strong intensity rise in the August 27 event, which 
was accompanied by a sharp increase and decay in hardness, 
arrived as a single wave at the time $\T=\T_0+0.5$~s.
Thereafter the spectral variations became, as in the March~5 burst, quite 
moderate, with the average value $kT\simeq 25$~keV typical also 
of the recurrent bursts in SGR~1900+14 remaining constant.

On the whole, the information collected on the two superintense 
outbursts is not at odds with the assumption that the processes 
accounting for emission of the narrow initial pulse and the long 
pulsating tail are separated in the source not only in time but 
in space as well, as is proposed in the model of Thompson and 
Duncan~(1995).

The characteristics of the two bursts are listed 
in Table~1. Table~2 contains the burst energetics in the 
sources. The obvious similarity between the bursts suggests an 
intimate similarity between the processes involved in their 
generation.

As follows from the above data, an explosive release of energy 
which is enormous even for a neutron star did not produce 
noticeable changes in the characteristics of SGR~1900+14. We have 
already pointed out that the properties of recurrent bursts 
remained practically unchanged (Mazets et al., 1999).
No substantial variations in the 
neutron-star rotation period which could be associated with the 
outburst were observed (Kouveliotou et al., 1999). At the same time the profile of the 
5.16-s pulsations in the weak persistent X-ray flux did undergo 
changes. Indeed, the multi-peak pulsations seen in May 
1998 were replaced, judging from the 1998 August~28 observations, 
by a single-peak pattern (Kouveliotou et al., 1999). This transition 
may have been very fast, during the pulsation stage of the 
outburst (see Fig.~7).

Detection of two giant outbursts in different sources during 
$t=20$~years undoubtedly implies that such bursts must be 
recurrent events. As pointed out more than once (Kouveliotou et al., 1994;
Norris et al., 1991, Hurley et al., 1994), the number $N$ of neutron stars 
in the Galaxy residing at any one time in the SGR stage, should be small, 
$\sim 7$. Accordingly, the average time interval $\tau$ between giant 
outbursts in the same repeater should not be large. Assuming $k$ 
bursts to occur in $N$ sources in $t$ years, rough estimates made 
by the maximum likelihood method yield $\tau \sim Nt/k \sim 50-100$~years.

It should be pointed out that the sensitivity of present-day 
$\gamma$-ray burst detectors is high enough to permit observation of 
the initial pulses in giant outbursts from soft gamma repeaters 
in the Local-Group of galaxies, primarily in the Andromeda Galaxy. For 
the distance to such sources of $\sim 700$~kpc, the bursts should 
appear as short, $\sim 0.2$-s long, spikes with an intensity
$\sim 10^{-6}$~erg~cm$^{-2}$. Their occurrence frequency should 
be of the order of one event per decade. Given a certain luck, such 
bursts could be detected and localized.

A neutron star in the stage of active soft gamma repeater expends 
its energy in supporting emission of a soft X-ray source, giant 
outbursts, and weaker recurrent bursts. To maintain such an 
activity during $\sim 10^4$~years, the initial energy store in the 
source should be at least $10^{47}$~erg. 

\acknowledgments
The authors thank G. M. Gorodinskii, A. A. Kolchin, 
and V. V. Lebedev for their assistance with the laboratory calibration of 
the 
equipment. 

Support of the Russian Space Agency and Russian Basic Research 
Foundation (Grant 99-02-17031) is gratefully acknowledged.

\begin{deluxetable}{lcc}
\tablecaption{Comparative characteristics of the giant outbursts}
\tablehead{
\colhead{}& \colhead{\bf{SGR~1900+14}}& \colhead{\bf{SGR~0526-66}}\\
\colhead{}& \colhead{$E_\gamma > 15$~keV}& \colhead{$E_\gamma > 30$~keV}
}    

\clearpage

\startdata
\bf{Giant outburst}& 			August~27, 1998& 	March~5, 1979\\
\tableline
\bf{Precursor}&				$\T-\T_0 = -0.45$~s&	?\\
&					$kT\sim 20$~keV&	\\
\tableline
\bf{Initial pulse}&			&			\\
Duration&				$\sim 0.35$~s&		$\sim 0.25$~s\\
Steep rise time&			$< 4$~ms&		$< 2$~ms\\
Exponential falloff&			$\tau _1\sim 35$~ms&	$\tau _1\sim 40$~ms\\
Peak flux F, $erg~cm^{-2}~s^{-1}$&	$>3.1\times 10^{-2}$&	$1\times 10^{-3}$\\
Fluence S, $erg~cm^{-2}$&		$>5.5\times 10^{-3}$&	$4.5\times 10^{-4}$\\
Spectral parameter $kT$, evolution&	$300 - 20$~keV&		$\sim 500$~keV\\

\tableline
\bf{Single wave}&			&			\\
Time interval, $\T-\T_0$&			$\sim 0.35-0.8$~s&	$\sim 0.25-1.5$~s\\
Ocsillations, quasiperiod P&		$\sim 0.08$~s&		$\sim 0.15$~s\\
Peak flux F, $erg~cm^{-2}~s^{-1}$&	$\sim 1.5\times 10^{-3}$& $\sim 3\times 10^{-5}$\\
Fluence S, $erg~cm^{-2}$&		$\sim 3\times 10^{-4}$&	$\sim 4\times 10^{-5}$\\
Spectral parameter $kT$, evolution&	$20 - 250$~keV&		$\sim 30$~keV\\
\tableline
\bf{Tail}&				&			\\
Exponential decay&			$\tau _2\sim 90$~s&	$\tau _2\sim 100$~s\\
Period $P$&				$5.16$~s&		$8.0$~s\\
Fluence S, $erg~cm^{-2}$&		$4.2\times 10^{-3}$&	$1\times 10^{-3}$\\
Spectral parameter $kT$&			$\sim 20$~keV&		$\sim 30$~keV\\
\tableline
\bf{Recurrent bursts}& 			& 			\\
Observations& 				May~1998 - January~1999& March~1979 - April~1983\\
Duration&				$\sim 0.1\div 4$~s&	$\sim 0.1\div 9$~s\\
Peak flux F, $erg~cm^{-2}~s^{-1}$&	$2\times 10^{-6}\div 3\times 10^{-5}$& 	$1\times 10^{-6}\div 7\times 10^{-6}$\\
Fluence S, $erg~cm^{-2}$&		$2\times 10^{-7}\div 5\times 10^{-5}$& 	$1.5\times 10^{-7}\div 2\times 10^{-5}$\\
Spectral parameter $kT$&			$20\div 30$~keV&		$30\div 35$~keV\\
\enddata
\end{deluxetable}

\begin{deluxetable}{lcc}
\tablecaption{Luminosity and energy release in SGR~1900+14 and SGR~0526-66}
\tablehead{
\colhead{}& \colhead{\bf{SGR~1900+14}}& \colhead{\bf{SGR~0526-66}}\\
\colhead{}& \colhead{$E_\gamma > 15$~keV}& \colhead{$E_\gamma > 30$~keV}
}    

\startdata
\bf{Distance}&				$10$~kpc~$^{(a)}$&		$55$~kpc\\
\tableline
\bf{Giant outburst}& 			August~27,1998& 	March~5,1979\\

\tableline
\bf{Initial pulse}&			&			\\
Energy release Q, $erg$&		$>6.8\times 10^{43}$&	$1.6\times 10^{44}$\\
Peak luminosity L, $erg~s^{-1}$&	$>3.7\times 10^{44}$&	$3.6\times 10^{44}$\\

\tableline
\bf{Tail}&				&			\\
Energy release Q, $erg$&		$5.2\times 10^{43}$&	$3.6\times 10^{44}$\\

\tableline
\bf{Total energy release Q}, $erg$&	$>1.2\times 10^{44}$&	$5.2\times 10^{44}$\\

\tableline
\bf{Recurrent bursts}& 			& 			\\
Observations& 				May~1998 - January~1999& March~1979 - April~1983\\
Energy release Q, $erg$&		$2\times 10^{39}\div 6\times 10^{41}$&	$5\times 10^{40}\div 7\times 10^{42}$\\
Peak luminosity L, $erg~s^{-1}$&	$2\times 10^{40}\div 4\times 10^{41}$&	$3\times 10^{41}\div 3\times 10^{42}$\\

\tableline
\bf{X-Ray source}& 			& 			\\
Luminosity L, $erg~s^{-1}$&	$\sim 10^{36}~^{(b)}$,$\sim 10^{35}~^{(c)}$&	$\sim 10^{35}~^{(d)}$\\

\tableline
\tableline
$^{(a)}$ Case, Bhattacharya, 1998&&\\
$^{(b)}$ Kouveliotou et al., 1999&&\\
$^{(c)}$ Hurley et al., 1999b&&\\
$^{(d)}$ Rothschild, Kulkarni, Lingenfelter, 1994&&\\

\enddata
\end{deluxetable}
\clearpage

\section*{FIGURE CAPTIONS}

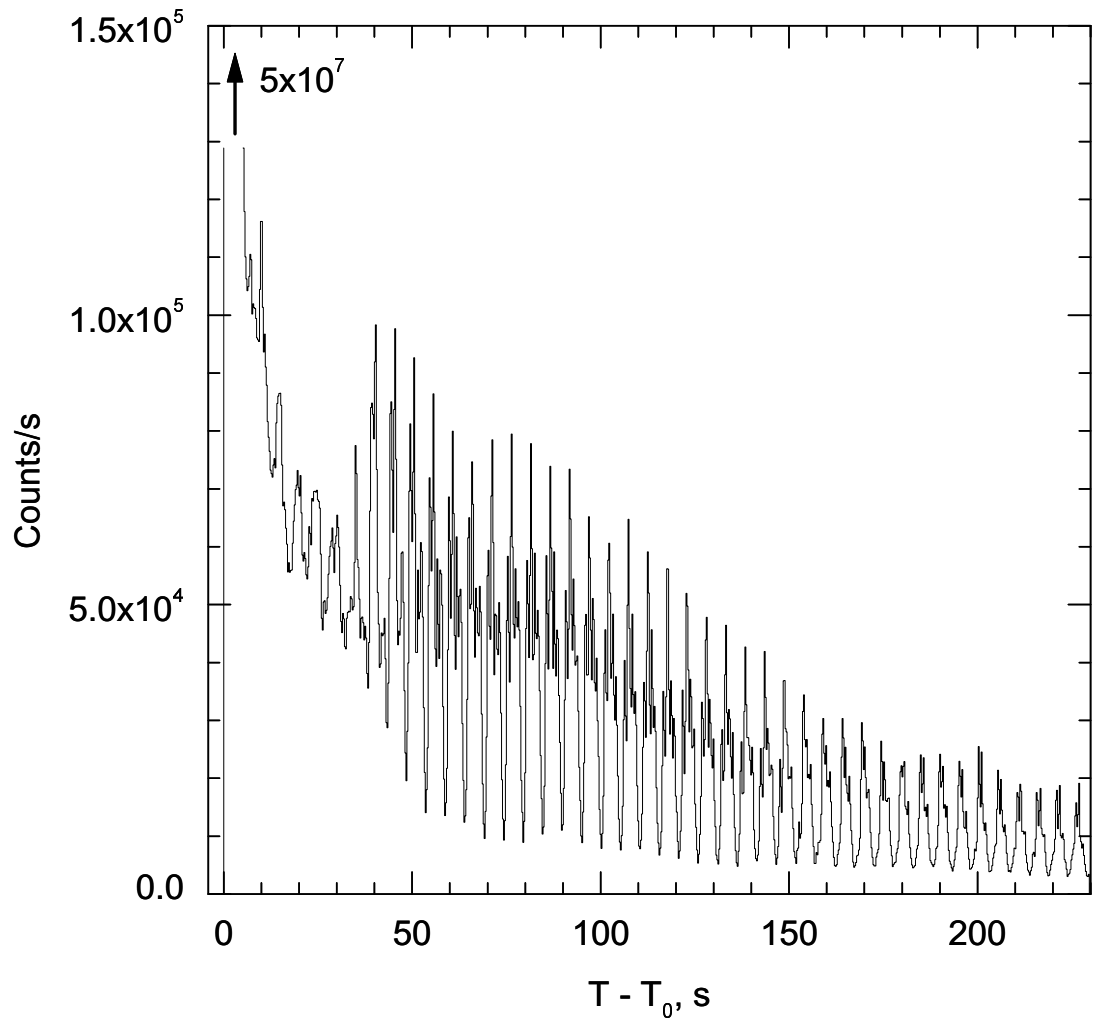
\figcaption{The giant 1998 August 27 outburst. Intensity of the 
$E_\gamma>15$~keV radiation.}

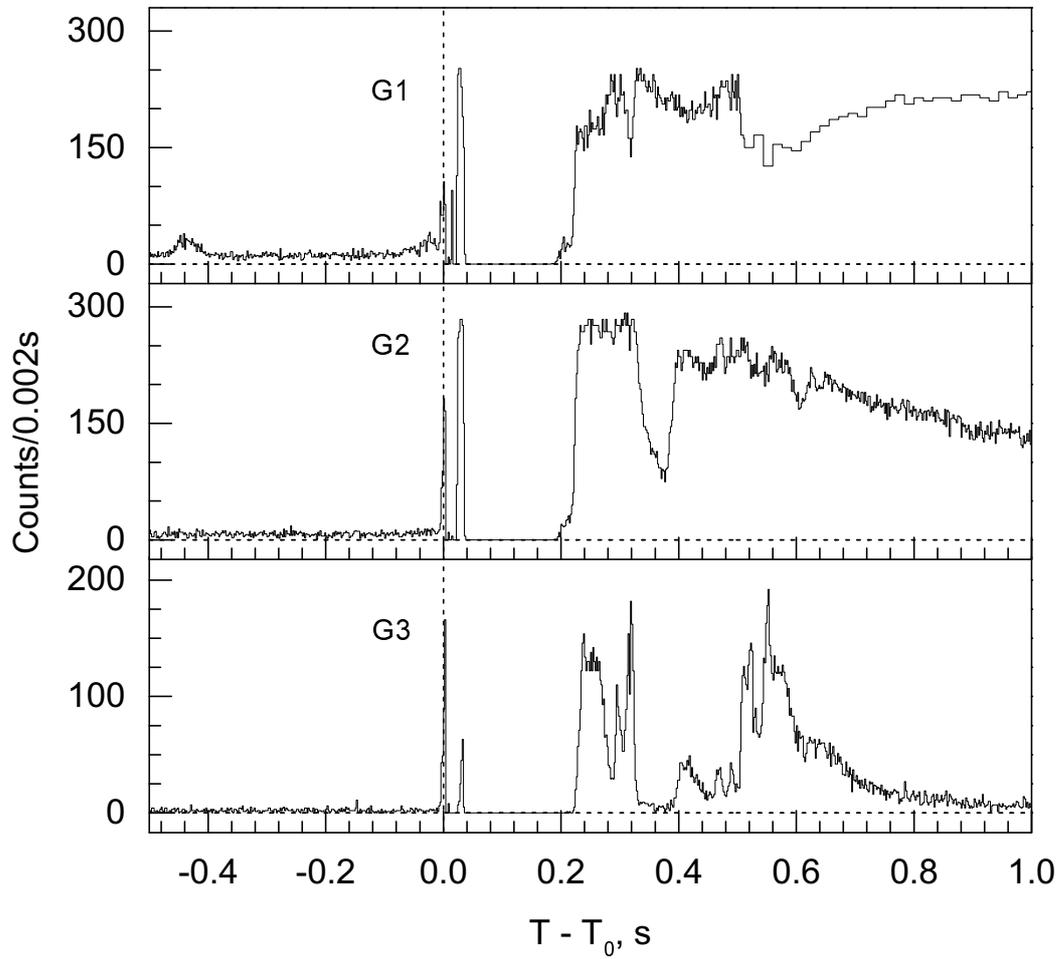
\figcaption{The initial phase of the burst recorded in three energy 
windows. Total detector overload for $\T-\T_0<0.2$ s.}

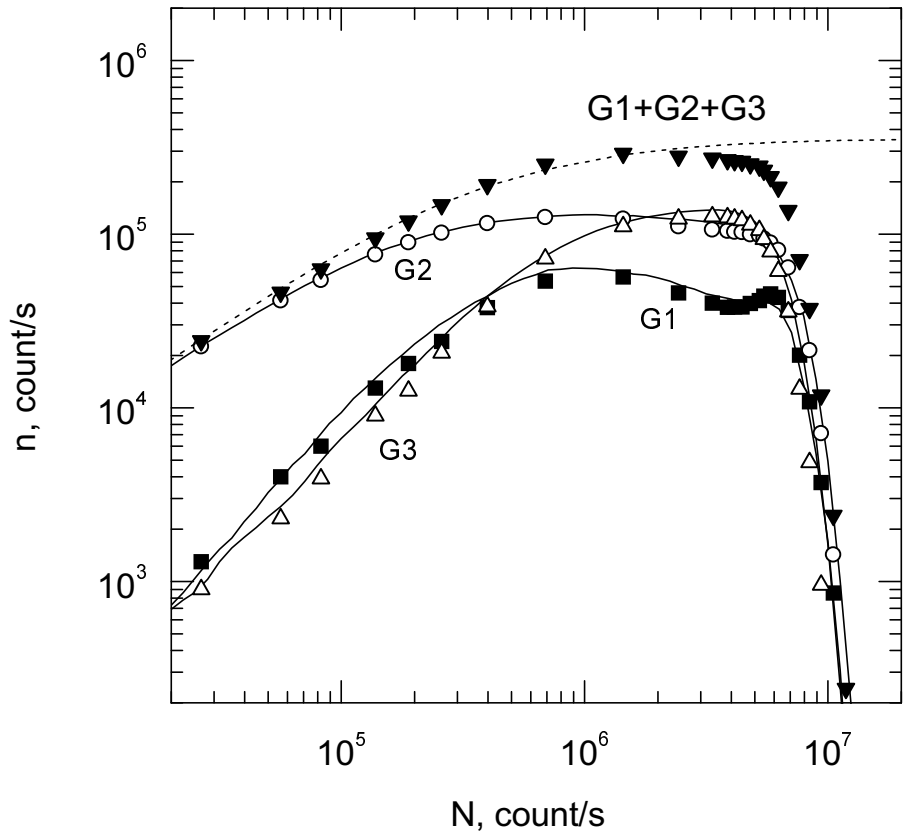
\figcaption{Count rates in three energy windows G1, G2, and G3 as 
functions of load (irradiation with a broad line with an energy 
of $\sim 100$ keV). Symbols -- laboratory measurements. Solid 
lines -- numerical simulation. Dashed line -- a plot of the standard relation 
$n(N,\tau)$ for $\tau=2.85\times 10^{-6}$~s.}

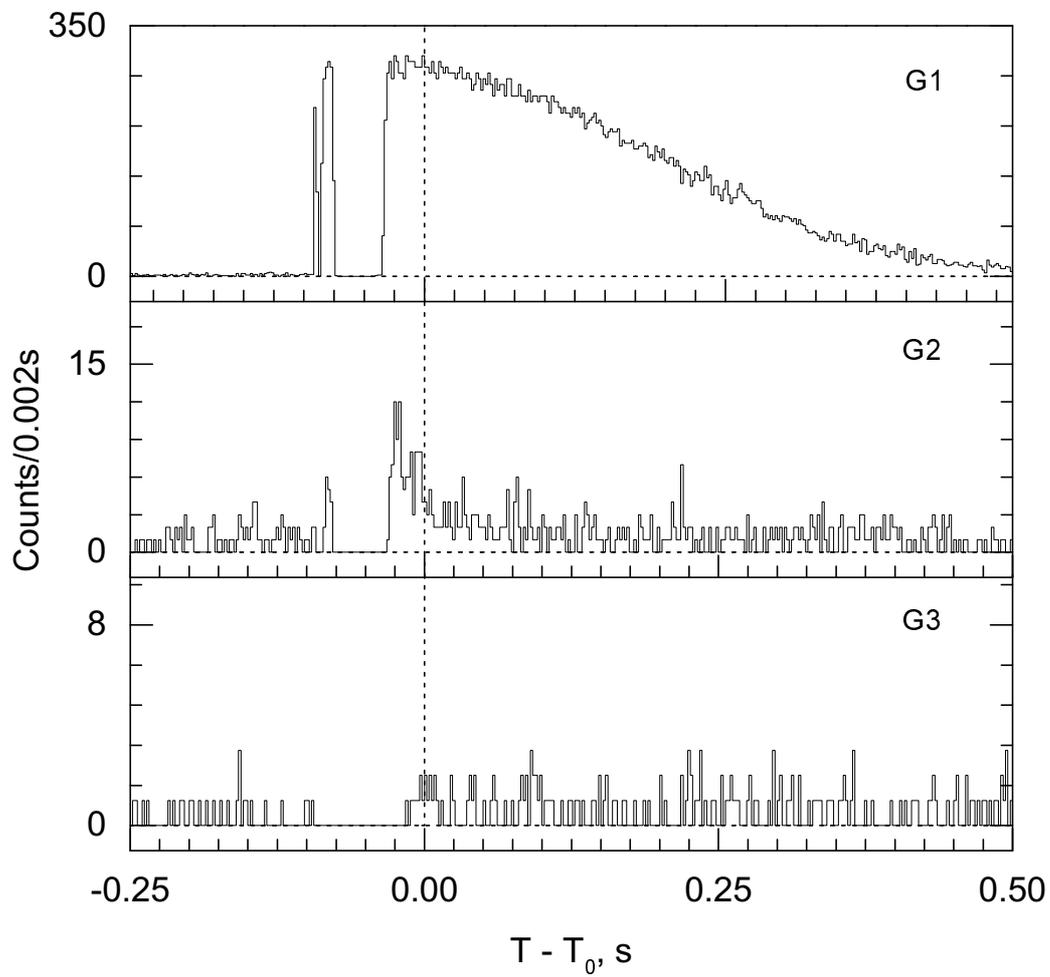
\figcaption{Imitation of a short burst caused by a nuclear 
interaction in the NaI(Tl) scintillator with the release of 
enormous energy $\sim 2.5\times 10^3$~GeV. The characteristic 
features of the overload are similar to those in the initial 
phase of the August 27 event.}

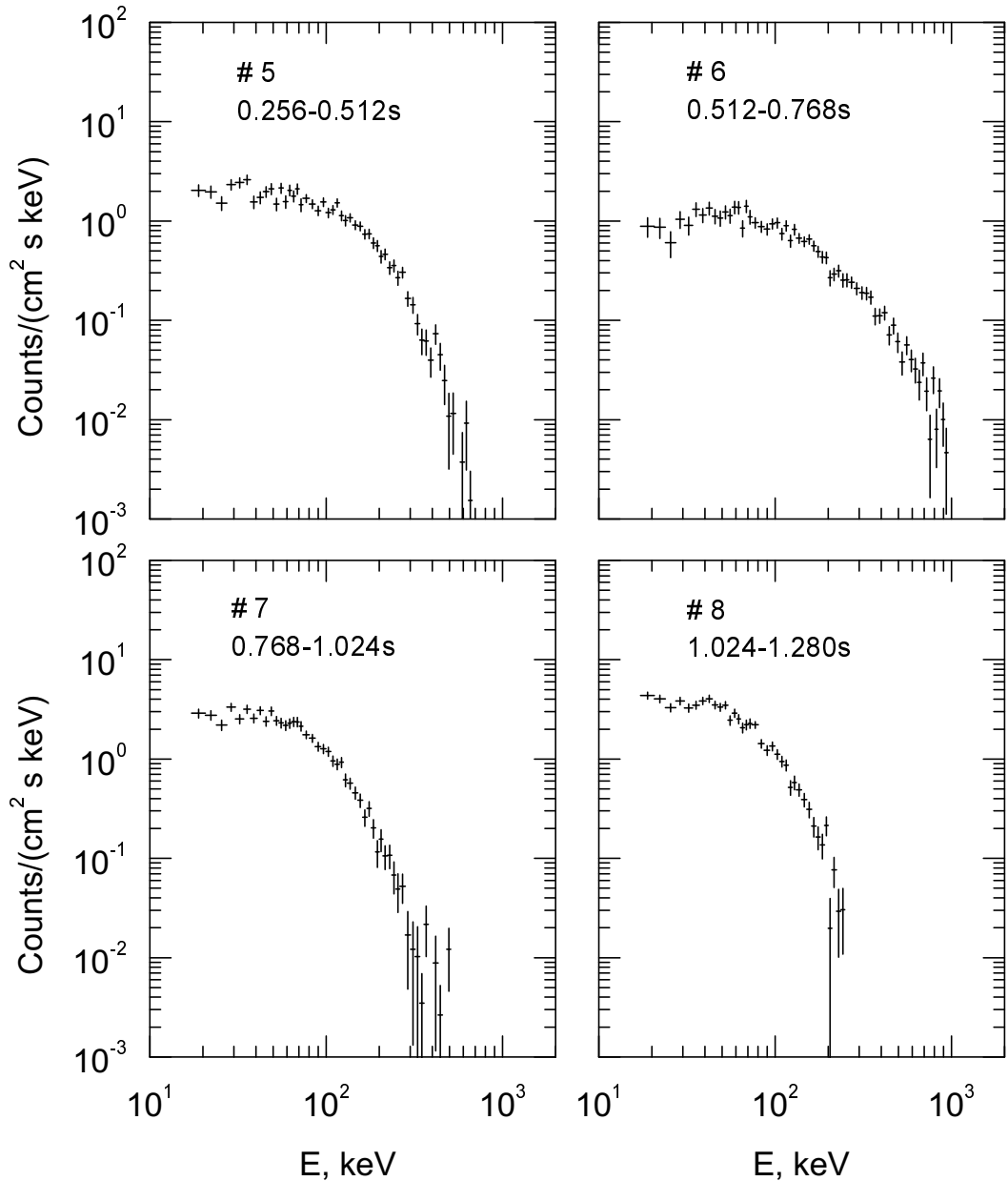
\figcaption{Energy loss spectra for the interval $\T-\T_0=0.256-1.280$~s
display extremely strong distortions of the incident 
photon spectrum because of pulse superposition. The incident 
spectrum within the interval $\T-\T_0=0.512-0.768$~s is obviously 
substantially harder.} 

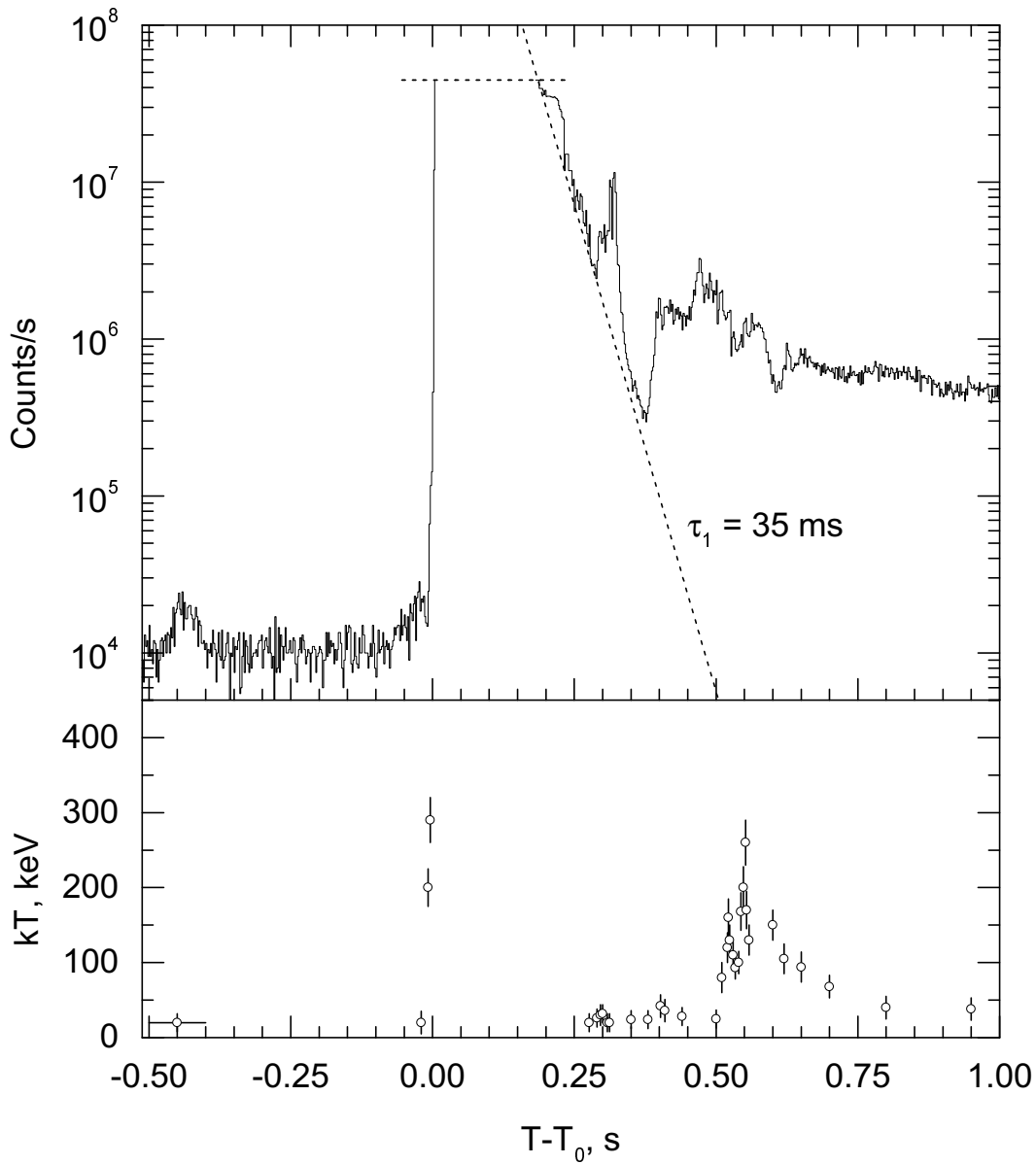
\figcaption{Reconstructed time history for the first second of the 
burst: radiation intensity with $E_\gamma>15$ keV and calculated 
parameter $kT$. The horizontal dashed line specifies the low 
intensity threshold causing total overload. The sloped line is a 
plot of the relation $\exp(-t/\tau)$ for $\tau_1=35$~ms.}

\noindent
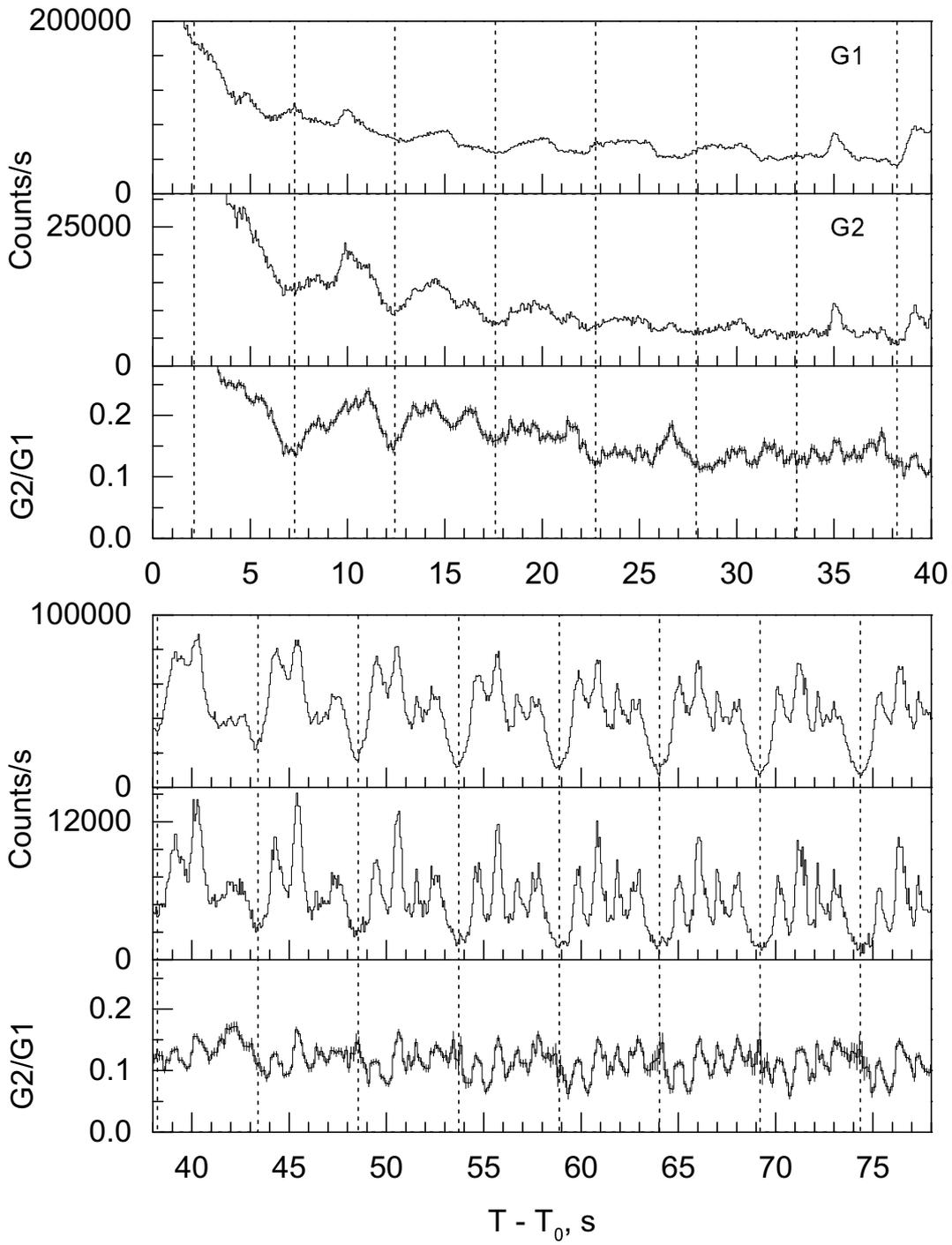
Fig. 7a.--- Time profile of the pulsating stage.
           The vertical dashed lines are spaced by 
            the pulsation period of 5.16~s.

\noindent
Fig. 7b.--- Time profile of the pulsating stage (continuation).

\noindent
Fig. 7c.--- Time profile of the pulsating stage (continuation).

\noindent
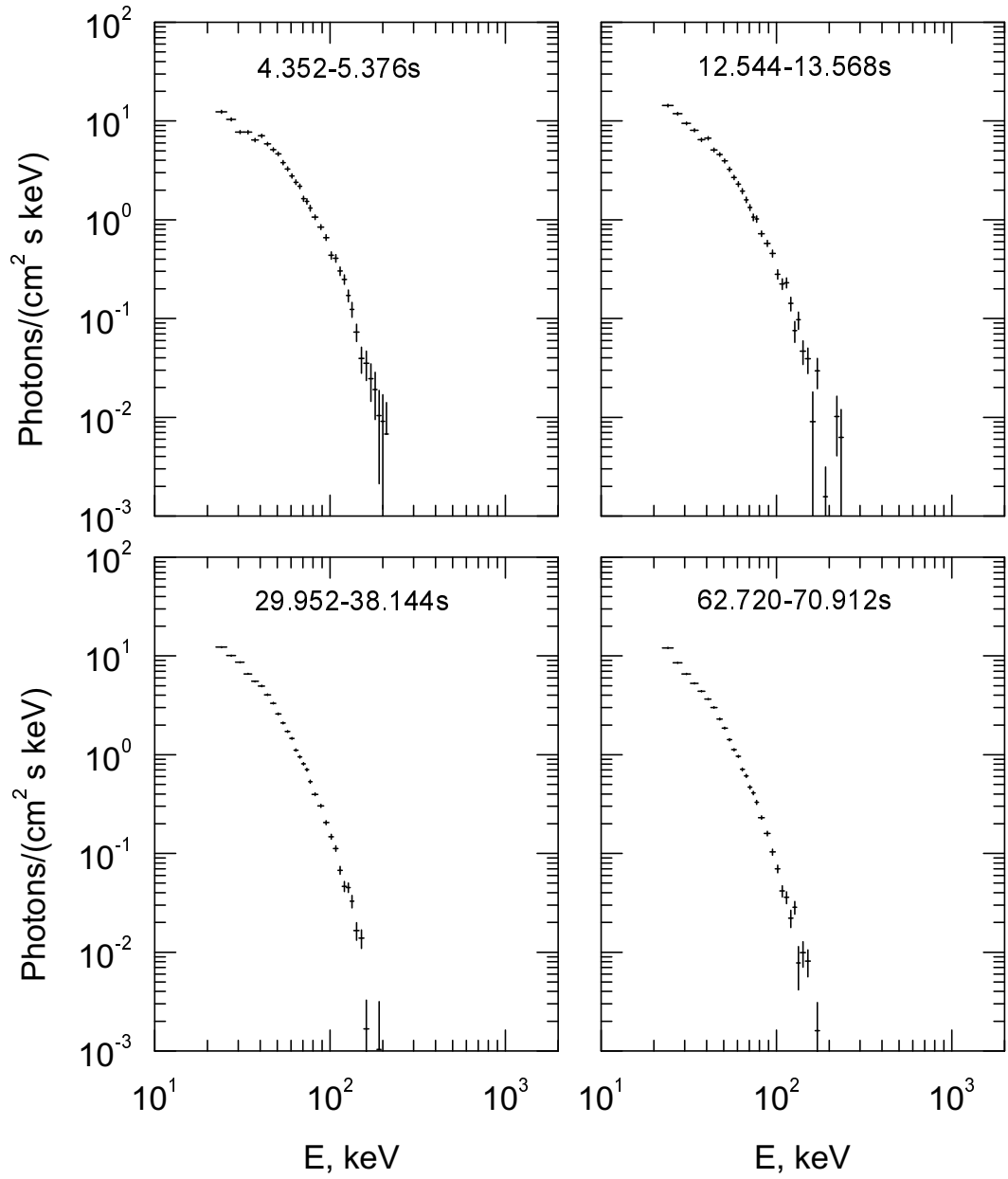
Fig. 8.--- Energy spectra of the burst pulsating stage.

\noindent
Fig. 9.--- Time and energy characteristics of the March 5 event.

\noindent
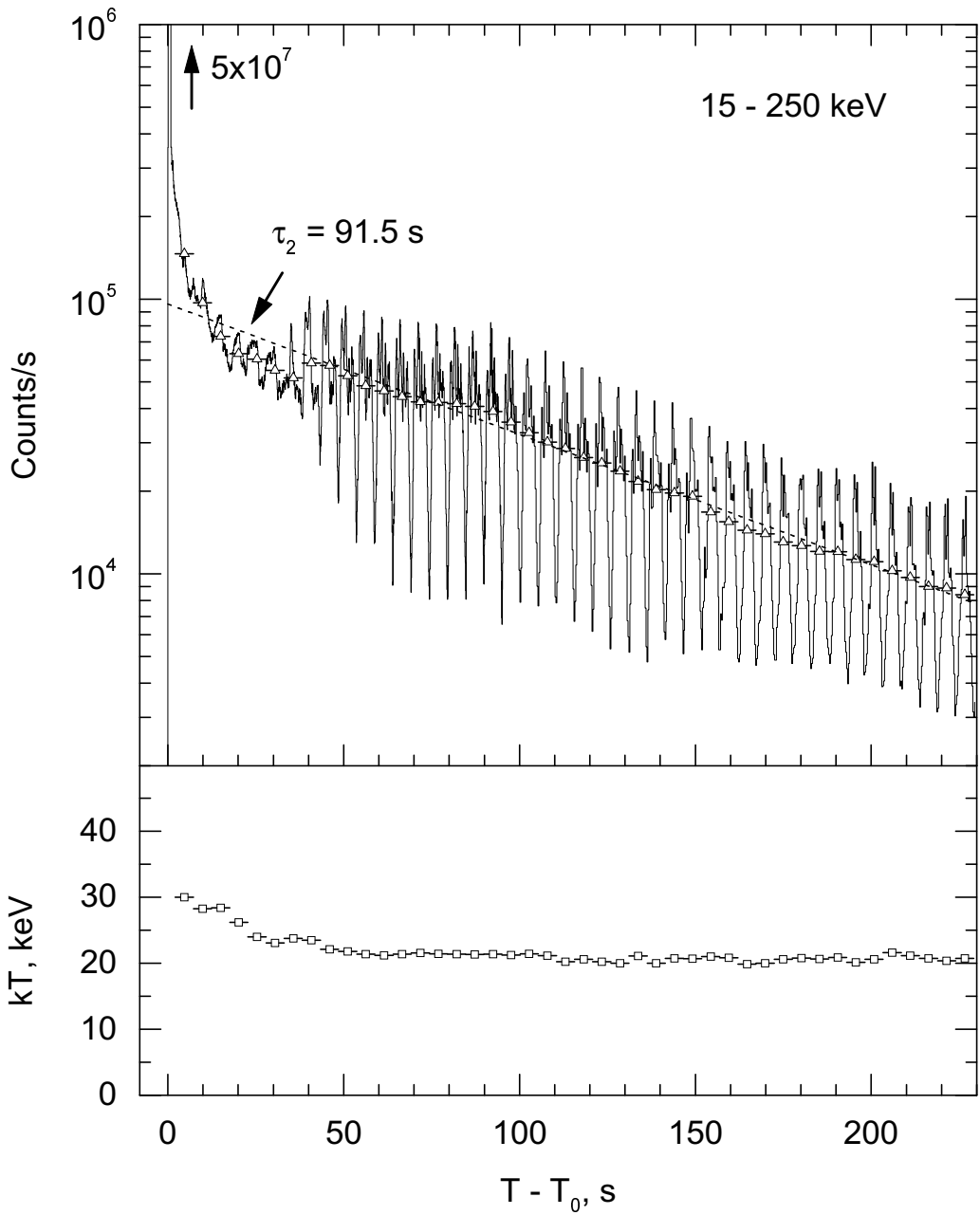
Fig. 10.--- Time and energy characteristics of the August 27 event.


\clearpage

\begin{references}

\reference{} Aptekar, R.L., et al. 1995, Space Science Rev., 71, 265
%
\reference{} Band, D., et al. 1993, ApJ, 413, 281
%
\reference{} Case, G.L., Bhattacharya, D. 1998, ApJ, 504, 761
%
\reference{} Cline, T.L., Mazets, E.P., \& Golenetskii, S.V. 1998, IAU Circ. 7002
%
\reference{} Fenimore, F.E., Klebesadel, R.W., \& Laros, J.G. 1996, ApJ, 460, 964
%
\reference{} Feroci, M., et al. 1999, ApJ, 515, L9
%
\reference{} Golenetskii, S.V., Ilyinskii, V.N., \& Mazets, E.P. 1984, Nature, 307, 41
%
\reference{} Hurley, K., et al. 1994, ApJ, 423, 709
%
\reference{} Hurley, K., et al. 1999a, Nature, 397, 41
%
\reference{} Hurley, K., et al. 1999b, ApJ, 510, L111
%
\reference{} Koi\v{c}ki, S., Koi\v{c}ki, A., Ajda\v{c}i\'c., V. 1973, 
Nuclear Instruments and Methods, 108, 297  
%
\reference{} Kouveliotou, C., et al. 1994, Nature, 368, 125
%
\reference{} Kouveliotou, C., et al. 1999, ApJ, 510, L115
%
\reference{} Mazets, E.P., et al. 1979a, Nature, 282, 587
%
\reference{} Mazets, E.P., Golenetskii, S.V., \& Guryan, Yu.A. 1979b, Soviet. Astron. Lett., 5(No6), 343
%
\reference{} Mazets, E.P., et al. 1982, Ap\&SS, 84, 173
%
\reference{} Mazets, E.P., et al. 1999, Astronomy Letters, in press
%
\reference{} Norris, J.P., Hertz, P., \& Wood, K.S. 1991, ApJ, 366, 240
%
\reference{} Rothschild, R., Kulkarni, S., \& Lingenfelter R. 1994, Nature, 368, 432
%
\reference{} Thompson, C. \& Duncan, R.C. 1995, MNRAS, 275, 255
%
\end{references}
\end{document}